\documentclass[aps,prl]{revtex4}
\def\ee{\end{equation}}
\def\bea{\begin{eqnarray}}

\def\bra#1{\langle #1 |}
\def\ket#1{| #1\rangle}

\def\braket#1#2{\langle \, #1 \, | \, #2 \, \rangle}
\def\braopket#1#2#3{\langle \, #1 \, | \, #2 \, | \, #3 \rangle}
\def\Tr{{\rm Tr}}

\def\Prob{{\rm Prob}}
\begin{document}

\title{A Solution to the Lorentzian Quantum Reality Problem}

\author{Adrian \surname{Kent}}
\affiliation{Centre for Quantum Information and Foundations, DAMTP, Centre for
  Mathematical Sciences, University of Cambridge, Wilberforce Road,
  Cambridge, CB3 0WA, U.K.}
\affiliation{Perimeter Institute for Theoretical Physics, 31 Caroline Street North, Waterloo, ON N2L 2Y5, Canada.}
\email{A.P.A.Kent@damtp.cam.ac.uk} 

\date{September 2013 (revised March 2014)} 

\begin{abstract}
The quantum reality problem is that of finding a mathematically 
precise definition of a sample space of configurations of
beables, events, histories, paths, or other mathematical objects, 
and a corresponding probability distribution, for any given closed 
quantum system.  Given a solution, we can postulate that
physical reality is described by one randomly chosen configuration
drawn from the sample space.     
For a physically sensible solution, this postulate should imply
quasiclassical physics in realistic models.  
In particular, it 
should imply the validity of Copenhagen quantum theory and 
classical dynamics in their respective domains.
A Lorentzian solution applies to 
relativistic quantum theory or quantum field theory in Minkowski space
and is defined in a way that respects Lorentz symmetry. 
We outline a new solution to the non-relativistic and Lorentzian
quantum reality problems, and associated 
new generalizations of quantum theory.   
\end{abstract}
\maketitle
  
\section{Introduction}

Quantum theory is a mathematically beautiful theory that 
unifies all of known physics with the exception of gravity.  
Its probabilistic predictions for experimental outcomes have 
been verified for a very large range of physical phenomena,
and contradicted by no experiment.    
Yet, as John Bell so eloquently and persuasively 
argued \cite{bell2004speakable}, we 
do not know what precisely it is that quantum probabilities 
are probabilities of.   We do not have a mathematically precise
description of what Bell called \cite{bell1976theory,bell1987beables}
the ``beables'' for quantum theory --
a sample space of events, or histories, or paths,
or other mathematical objects, on which the quantum probability
distribution is defined.   This is the quantum reality problem, 
sometimes referred to as the measurement problem -- rather misleadingly
from a modern perspective, since few physicists now believe that
the fundamental laws of nature involve measuring devices {\it per
  se} or that progress can be made by analysing them.
As Bell emphasized, the quantum reality problem becomes
particularly conceptually 
problematic when we impose the natural condition that any
solution should respect the symmetries of special relativity.
We focus here on solutions to the
{\it Lorentzian quantum reality problem}, i.e. solutions 
that have this property. 

As Bell also stressed, \cite{bell2004speakable}, mathematical aesthetics are not
the main motivation for solving the quantum reality problem.
The motivation is the following.  On the one hand, the impressive
successes of quantum theory, and the lack of compelling alternatives,
make it natural to try to treat quantum theory as fundamental, and
so to derive everything else in physics from quantum theory.
On the other hand, it appears to us that we live in a 
quasiclassical world, in which macroscopic variables are
most of the time approximately governed by deterministic
equations of motion, but are also affected by random
events of quantum origin.  Moreover, it appears as though
this quasiclassical world emerged from an initial quantum
state with no initial quasiclassical properties. 
Given a well-defined probabilistic version of the quantum theory
of closed systems, we can hope to explain these features from
within quantum theory, and indeed to sketch a coherent and unified account of 
cosmology, classical and quasiclassical dynamics and quantum theory.
Without one, we cannot rigorously derive classical or quasiclassical physics 
from quantum theory, nor give a coherent treatment of cosmology
from within quantum theory.   

The once-standard Copenhagen
interpretation of quantum theory explicitly accepted these limitations.
It is the hope of going further, and giving a unified framework 
that includes all of modern physics, that motivates the ongoing 
search for a solution.    

The first 
well-known attempt to address the quantum reality problem directly was the pilot
wave theory of de Broglie  and Bohm  \cite{debroglie, bohm}, 
in which the beables are 
particle trajectories whose evolution is defined by the quantum
wave function by a guidance equation.  However, de Broglie and Bohm's
models apply to non-relativistic quantum mechanics and are
inconsistent with special relativity.  No
fundamentally relativistic generalisation of the models has been
found, nor is there a convincing extension to quantum field theory.  
Many (though not all) 
physicists also find de Broglie and Bohm's trajectories and
guidance equations rather mathematically unnatural and inelegant 
additions to quantum theory. 

Non-relativistic dynamical collapse models
\cite{ghirardi1986unified,GPR} 
attempt to give another
story about physical reality that is consistent with experiment
to date, at the price of changing the dynamics and hence the experimental
predictions of quantum theory.   (For some attempts in the 
direction of relativistic collapse models 
see \cite{pearle1999relativistic,pearle2005quasirelativistic,
tumulka2006relativistic,bedingham2011relativistic}.)  
While scientifically interesting, these and other
generalizations of quantum theory do not address the main question 
we focus on here -- namely, 
whether we can find a mathematically precise description of 
reality consistent with standard quantum theory.     

Another line of thought, initiated by Everett, suggests that 
quantum theory is deterministic and that pure unitary quantum
evolution holds at all times.  The problems with this idea,
and with the many incompatible
proposals for some form of ``many worlds'' quantum theory
that it has inspired, continue to be debated \cite{mwbook}.  
Still, two relatively uncontroversial points can be made.  
First, since, according to most modern Everettians, 
quantum theory is fundamentally
deterministic, and the appearance of quasiclassical physics is
supposed to arise as an approximation
via decoherence, no mathematically precise sample space and probability
distribution emerges.  
Second, many worlds theories are radically different types of
scientific theory from standard ``one world'' versions of quantum
theory (or indeed from anything previously considered in science)
and give a qualitatively different (and fantastically weird)
description of reality. 

Many critics
are also unpersuaded either that the appearance of 
quasiclassical physics can in fact be explained by 
decoherence, or that a fundamentally deterministic theory
can account for an approximate higher level description that 
explains the empirical appearance of Born rule probabilities
(see e.g. \cite{kentoneworld,albert,price}).  
But even convinced Everettians, 
who believe that many worlds theories can 
give a complete explanation of 
the Born rule, the appearance of
probabilities and of quasiclassicality, and everything else 
described by standard physics, should still be (and generally
are) interested in whether we {\it need} to invoke many worlds ideas
to do all this.  

In short, there remains an unresolved and intellectually fascinating question 
fundamental to our understanding of quantum theory.
Namely, is there a mathematically precise solution to the 
quantum reality problem, consistent with the symmetries
of special relativity, that gives a probabilistic 
description of one physical world, consistent with
the quasiclassical combination of classical and quantum physics
that we actually observe?  Or, if this is too much to presently
hope, given that any fully realistic model of quasiclassical physics
would need to describe the real time evolution of 
complex physical structures within quantum field theory and 
quantum gravity, is there at least a conceptually clear route
to defining such a solution?  

This paper answers this last question positively.  
The solution described here uses 
the strategy of inferring finite time beables from asymptotic
behaviour and many of the other ideas
set out in Ref. \cite{akrealworld}, but is
simpler than the proposals made in that paper. 

\section{The reality problem for 
nonrelativistic quantum mechanics} 

We should note at the start that the intuitions underlying 
our proposed solution come from the properties of relativistic
quantum field theories and from the current tentative understanding
of the likely asymptotic state of the universe within presently
favoured cosmological models.  Thus, although the ideas are described
more simply in nonrelativistic quantum mechanics than in 
relativistic theories, nonrelativistic quantum mechanics is not
ultimately the most natural setting for them, and our solution will
not necessarily give intuitively appealing descriptions of
reality for simple models of nonrelativistic systems, without
further assumptions.   
We discuss further the underlying intuitions, and the
assumptions that need to be introduced into models to 
reflect them, below, after giving the proposal in general form.  

Suppose that we have a system of $N$ particles, including 
$b$ indistinguishable bosons and $f$ indistinguishable fermions,
with $b+f  \leq N$, and $(N-b-f)$ distinguishable particles. 
(These choices are made purely 
to simplify our discussion, which can easily
be extended to allow $M_b$ types of bosons and $M_f$ types of fermion, 
for any integers $M_b , M_f \geq 0$.) 
We take $b,f >1$ (otherwise we treat the relevant particle
as distinguishable) and label the bosons by $ \{ 1 , \ldots , b \}$
and the fermions by $\{ b+1 , \ldots , b+f \}$; if $N - 
b - f > 0$ we label the remaining distinguishable
particles by $\{ b+f+1 , \ldots , N \}$.
We also suppose that the $N$ particles have some natural division
into two classes, which we label $1$ and $2$.  All 
indistinguishable particles of the same type belong to the same
class.  Both classes contain significant numbers of particles, and
there are significant interactions between the particles in class
$1$ and those in class $2$, which allow the final states of the class
$1$ and class $2$ particles to be highly correlated.  

Our proposal involves (as mathematical abstractions) final and 
intermediate time measurements of the masses at points in space, which
are functions of particle position operators.   Position (more
precisely, the mass at a given position) thus plays
a special role as a mathematically preferred observable.     
We thus suppress spin and any other internal degrees of freedom to simplify
the notation; note that our proposal requires
no additional spin or other measurements to be introduced
in systems where they are relevant.   

The system's position space wave function has the appropriate
statistics. 
If $\rho_b$ is a permutation of  $ \{ 1 , \ldots , b \}$
and $\rho_f$ is a permutation of  $\{ b+1 , \ldots , b+f \}$, 
write $\rho = \rho_b \otimes \rho_f \otimes I_{N - b - f}$ 
for the corresponding permutation of $\{1 , \ldots
, N \}$.  Then we have 
$$
\psi (x_{\rho(1)} , \ldots , x_{\rho(N)} ) = \epsilon ( \rho_f ) 
\psi (x_{1} , \ldots , x_{N} ) \, , 
$$
where $\epsilon$ is the sign of $\rho_f$.  

We will treat this as a closed system without external intervention,
which comes into existence at $t=0$ and continues to a final time
$t=T$, at which point time and physics end.   This is a mathematical
device, not a fundamental assumption about nature.   
Later we will consider the limit $T \rightarrow \infty$, which gives
a more conventional (although in this model still non-relativistic)  
picture in which physics begins at some point in the
past and continues forever thereafter.  

Suppose we are given the initial state $\ket{\psi(0)}$ at $t=0$ 
with wave function 
$$
\psi (x_{1} , \ldots , x_{N} ; 0 ) = \braket{x_1 , \ldots ,
  x_N}{\psi(0)} \, ,  
$$
and that we are given a Hamiltonian $H$.
Quantum theory then gives the Schr\"odinger evolution
$$
\ket{\psi(t)} = \exp ( -iHt / \hbar ) \ket{\psi (0 )} \, , 
$$
or in terms of wave functions
$$
\psi (x_{1} , \ldots , x_{N} ; t ) = \exp ( - i H t / \hbar ) \psi
(x_{1} , \ldots , x_{N} ; 0 )
\, . 
$$

Now consider, for each of classes $1$ and $2$, the
coresponding mass density function defined by a mass-weighted
sum of position operators  
$$
\rho^{(j)} (x;t) = \sum_{i \in C^{j}} m_i  \rho_i (x,t ) \, . 
$$
for $j=1,2$.  Here the set $C^{j}$ denotes the labels of all
particles in class $j$, and 
\begin{eqnarray}
\rho_i (x,t ) &=&  \int d x_1 \ldots d x_{i-1} d x_{i+1} \ldots dx_N |
\psi ( x_ 1 , \ldots x_{i-1}, x, x_{i+1}, \ldots , x_N ; t ) |^2
\nonumber \\
&=& \braopket{ \psi (t)}{  P_i^x }{ \psi (t) } \, , 
\end{eqnarray}
where
$$P_i^x = I_1 \otimes \cdots I_{i-1} \otimes \ket{x}_i \bra{x}_i
\otimes I_{i+1} \otimes \cdots \otimes I_N 
$$ 
is the formal projection operator onto the value 
$x_i = x$ of the $i$-th coordinate.   
Note that, while $P^x_i$ is not even formally
well defined if $i$ is a bosonic or
fermionic label, the sums $\sum_{i=1}^b m_i P^x_i $ and 
$\sum_{i=b+1}^{b+f} m_i P^x_i $ are.  (The identical bosons have equal
masses, $m_1 = m_2 = \ldots = m_b$; similarly $m_{b+1} = m_{b+2} = \ldots =
m_{b+f}$.) 

The operators $ \sum_i m_i P_i^x $ and $ \sum_i m_i P_i^{y} $ 
commute.  So, formally, we can consider the effect of a simultaneous
measurement of all such operators at time $t=T$.   This produces
a possible final time distribution 
$$
\rho_f (x,T) = \rho_f^{(1)} (x, T) + \rho_f^{(2)} (x,T) = 
\sum_i m_i \delta ( x - y_i ) \, , 
$$
randomly chosen from the sample space of all distributions 
with total mass $\sum_i m_i$, via the probability distribution defined
by $\ket{\psi(0)}, H$ and the Born rule.   This is the first
ingredient in our construction of a beable description of
non-relativistic 
quasiclassical reality.   We will take the randomly chosen 
pair $\rho_f^{(1)} (x,T)$ and $\rho_f^{(2)} (x,T)$
to define the ``real world'' that is chosen from among the
sample space of possible worlds that could arise given 
the initial state and Hamiltonian.  This final time measurement is 
not meant to be thought of as carried out by any
external system: we treat the $N$ particle universe as a closed
system with no external observers or devices.   It is simply a mathematical
operation that allows a precise description of the sample
space of possible worlds and a corresponding probability
distribution. 

Next we consider the expected value of $\rho^{(j)}(x,t)$ (for $0 < t < T$), 
given the initial state $\ket{\psi(0)}$ and Hamiltonian $H$, 
when we condition on the outcome of our final time measurements
producing the final distribution $\rho^{(\bar{j})}_f (x,T)$. 
Here $\bar{j}$ denotes the other class to $j$; thus we consider 
expected values of $\rho^{(1)} (x,t)$ given the final distribution
$\rho^{(2)} (x, T)$ and vice versa.  

Since we have a post-selected final outcome,
this expectation value depends on precisely which set of 
commuting operators are simultaneously measured. 
For each point in time $t$, 
we consider a simultaneous measurement of $\rho^{(\bar{j})}(x,t)$ at all
points $x \in R^3$. 

Formally, the relevant expectation value is then
given by the Aharonov-Bergmann-Lebowitz
rule \cite{aharonov1964time}, extended to the case where the
intermediate and final projection operators may be degenerate.
We write $\{ M_k^j : k \in K \}$ for the possible nonzero outcomes of measuring
the mass of particles in class $j$
at a point in our system, so that each $M_k^j = \sum_{l \in L}
m_l $ for some nonzero subset $L \subseteq C^{(j)} \subseteq
\{ 1 , \ldots , N \}$. 
Write $P_M^x$ for the projection onto the space of states with
mass $M$ at point $x$.   
Then we have 

\begin{equation}
\langle \rho^{(j)} (x,t) \rangle =  \sum_k M_k 
 \frac{ A^j (M_k , x, t, T)}{B^j (t,T)} \, , 
\end{equation}
where 
\begin{eqnarray}
A^j (M_k, x, t, T) &=&
\sum_{\{ k_1 , \ldots , k_{r} : k_i \in C^{(j)} \, \forall i {\rm~and~}
\sum_{i=1}^{r} M_{k_i} + M_k =
 M^{(j)} \} } 
\int dy_1 \ldots dy_{r} \\
&& \qquad  \Tr ( P_f \exp ( - i H (T-t) /
\hbar ) P_{M_k}^x \prod_{i=1}^{N-1} P_{M_{k_i}}^{y_i}
 \exp ( - i H t / \hbar ) P_0 \exp (  i H t /
\hbar ) P_{M_k}^x \prod_{i=1}^{N-1} P_{M_{k_i}}^{y_i} \exp (  i H (T-t) / \hbar )  )
\, , \nonumber 
\end{eqnarray}
and
\begin{eqnarray}
B^j (t,T) &=& \sum_{\{ k_1 , \ldots , k_{r+1} : k_i \in C^{(j)} \forall i
  {\rm~and~}
\sum_{i=1}^{r+1} M_{k_i} =
 M^{(j)} \} } 
\int dy_1 \ldots dy_{r+1}  \\
&& \qquad \Tr ( P_f \exp ( - i H (T-t) /
\hbar )  \prod_{i=1}^{r+1}  P_{M_i}^{y_i}  \exp ( - i H t / \hbar ) P_0 
\exp (  i H t / \hbar ) \prod_{i=1}^{r+1}  P_{M_i}^{y_i}  \exp (  i H (T-t) /
\hbar ) ) \, . \nonumber 
\end{eqnarray}
Here $P_0 = \ket{\psi (0 ) }\bra{ \psi(0)}$ and 
$P_{f}$ is the projection onto the space of states for which
the class $\bar{j}$ particles have final 
mass density $\rho_f^{\bar{j}} (x,T)$, and
$$
M^{(j)} = \sum_{l \in C^{(j)}} m_l \, .
$$

That is, our definition of an expectation value for the mass density
of class $j$ particles at intermediate times uses only the
postselected data from the other class, $\bar{j}$. 

That is, 
$$
\langle \rho^{(j)} (x,t) \rangle = \frac{ \sum_k M_k \sum {\rm wt}
({\rm ~all~outcomes~including~mass~} M_k {\rm~at~} x )}
{\sum  {\rm wt}
({\rm ~all~outcomes~})} \, , 
$$
where ${\rm wt}()$ denotes the pre- and post-selected probability 
weights used in the above expression, and the outcomes considered are
of simultaneous mass measurements of all the class $j$ particles
at all points $y \in R^3$. 
We could also include projections onto the zero mass eigenspaces
at all points other than $y_1 , \ldots y_{r-1} , y_r $ in the
denominator, and at all points 
other than $y_1 , \ldots y_{r-1} , x$ in the numerator.   However, these would
not change the expressions here, since we have a fixed number
of positive mass particles of total mass $\sum_{l \in C^{(j)}} m_l$.  


This is the second ingredient in our construction: given the 
final outcomes $\rho^{(j)}_f (x,T)$, we take the expressions
$$
\rho^{(j)}_T (x, t) = \langle \rho^{(j)} (x,t) \rangle $$
just calculated (using the post-selected final conditions
for the complementary class $\bar{j}$ to define the beable
for class $j$) to define 
the beables at position $x$ and time $t$ for a universe
in which physics runs from time $0$ to time $T$. 

The full set of beables describing 
reality for our first model, in which all physics 
takes place between times $0$ and $T$, is thus given by 
\begin{equation}\label{realitymass}
\{ \, \rho^{(j)}_T (x,t)  \, : \, 0 < t < T \, , \, x \in R^3
\, , j = 1,2 \, \} \, . 
\end{equation} 

To make further progress, we need a key assumption: quantum physics in 
our model universe involves non-trivial interactions between
the particles in the two classes at finite times, creating
effective records, but 
becomes, in a sense to be characterised
more precisely, asymptotically trivial as $t \rightarrow \infty$.  

A possible intuition that would support this assumption is that, while
initially the particles often are localized in the same region and
interact, eventually all
particles that can decaywill have decayed, all particles that
are capable of interacting with one another either do interact
or become more and more widely separated, non-gravitational 
interactions become rarer and rarer, and the asymptotic evolution
is effectively described by a free quantum field theory.   
This intuition relies on being able to think of the asymptotic 
physical state as composed of elementary particles, or at least as behaving
qualitatively as though it were. 
It is supported by some cosmological
scenarios that are presently taken seriously, for example (and most
cleanly) in ``big rip'' scenarios.    
To model something like this in the nonrelativistic
setting requires in particular that interactions between particles
in class $1$ and those in class $2$ are initially significant
but are switched off, or become negligible, at large times.   

A weaker intuition, still adequate to justify the assumption, is
that the outcome of a typical indeterministic quasiclassical event 
leaves an indelible asymptotic record in the mass densities of 
one or both classes (both being required if the event itself 
is quasiclassical with respect to variables defined by both classes).
That is, 
in principle, a measurement of the class mass densities at large final time $T$ 
allows one to infer all initially undetermined outcomes, and so 
the entire history of quasiclassical physics, which is encoded in the
inferred mass density distributions at times between $0$ and $T$.  

For example, suppose that the Schr\"odinger equation 
creates what is traditionally thought of as a measurement event at
time $t<T$.  That is, suppose that, around time $t$,
a quantum system interacts with an
apparatus whose pointer initially has a single approximately
localized position, and 
creates a superposition of two states corresponding to 
macroscopically separated approximately localized
pointer positions at time $t + \delta$.
Suppose also that the pointer comprises
particles in class $1$, and its environment contains particles
in class $2$ that interact with it. 
In a fairly general class of such models, the  position
degrees of freedom of many particles in the environment typically become coupled
to the pointer positions, and produce effective records (i.e. multiply
redundant subsystems that are persistently correlated with the 
original data) of those positions in the
environment.   The class $2$ 
mass density measurement at the final time $T$ distinguishes
different states of these records and so indirectly
measures the pointer position at times soon after $t+ \delta$,
whether or not the pointer itself remains intact or quasiclassical
indefinitely.   

Of course, the point of giving a precise definition of
reality in terms of beables is to go beyond intuition.   In our
models, a definite quasiclassical measurement event is ultimately
a higher level description, which can generally only be approximately
characterised in the quasiclassical theory based on the beables.
Such an event occurs if and only if it leaves effective records
in the final time mass density measurement(s) for the relevant
class(es).   Thus, a hypothetical
experiment successfully demonstrating interference between
two paths of a macroscopic object would not produce definite events 
associated with one or other path, since the interference 
implies near-perfect isolation of the beams from all other 
particles, and hence a typical final time mass density measurement
outcome will give almost no path information.  
     
Intuitions aside, then, the precise assumption we need to make is 
that the probability distribution for the possible configurations
of beables describing reality (\ref{realitymass})
within each fixed time interval ${[} 0 , t {]}$, for any $t < T$,
has a well-defined limit as $T \rightarrow \infty$. 
Given models in which this holds, we can then test whether 
typical beable distributions represent quasiclassical reality
appropriately \footnote{This can also be tested in finite $T$ models 
if we assume the required asymptotic behaviour.}. 

Our asymptotic assumption translates as follows.
Let $C^{(j)}_t$ (for $j=1,2$) be any coarse-grained subsets
of the sets of continuous
functions 
$$\{ \rho^{(j)} (x, t') : x \in R^3 , 0 \leq t' \leq t \}$$
obeying (in our non-relativistic model) the constraint
$$
\int d^3 x \rho^{(j)} (x, t' ) = \sum_{l \in C^{(j)}} m_l 
$$
for any $t'$. 
Let 
$$
\Prob_T ( C^{(1)}_t , C^{(2)}_t )
$$
be the joint 
probability that $\{ \rho^{(j)}_T (x, t') : x \in R^3 , 0 \leq t' \leq t
f\}$ belong to $C^{(j)}_t$ for $j=1$ and $2$, given our constructed 
probability density
function on the set of possible $\rho^{(j)}_T$. 
Then, assuming that 
$$
\Prob_{\infty} (C^{(1)}_t , C^{(2)}_t ) = 
\lim_{T \rightarrow \infty} \Prob_T ( C^{(1)}_t , C^{(2)}_t ) 
$$
exists, we define this expression to be the probability that reality up to time
$t$ is described
by a time-evolving mass distribution for class $j$ particles
belonging to $C^{(j)}_t$, for $j=1$ and $2$.   
This, together with the additivity of the probability measure
on finite disjoint measurable subsets of the sample space,
completes the definition of the probability distribution on
the beable configurations, i.e. on the possible descriptions
of reality, in this non-relativistic model.  

We will consider other non-relativistic beable models below. 
First, though, we discuss relativistic generalizations of 
the above model. 

\section{The reality problem for relativistic quantum theory} 

It is not presently possible to give a completely rigorous 
discussion of the reality problem for any 
physically relevant relativistic quantum field theory in Minkowski
space, because no version of relativistic quantum field theory is
well enough understood to allow quasiclassical equations to be rigorously 
derived from first principles. 
We cannot even give a fully mathematically rigorous
quantum field-theoretic description of any
realistic physical experiment -- of electrons passing from a source
through a two-slit region and registering at detectors, for example. 
Evidently, then, we cannot hope to prove  rigorously that a particular 
mathematical construction attached to such a description gives
a description of physical reality with any given desired property.    

However, we {\it can} aim to separate the conceptual issue posed
by the reality problem from the technical issues that prevent us
from carrying out complete calculations describing realistic experiments,
or other phenomena characterised by 
quasiclassical physics, in quantum field theory.  We can also 
hope to make it plausible that a proposed solution correctly
describes quasiclassical reality in realistic models. 
This is the strategy we follow here.   

We now suppose that the initial state $\ket{\psi_0}$ is given on some spacelike 
hypersurface $S_0$, and that some relativistic unitary evolution law is
given.  The Tomonaga-Schwinger formalism allows us to define formally
the evolved state $\ket{\psi_S}$ on any hypersurface $S$ in the future
of $S_0$ via a unitary operator $U_{S_0 S}$.   
These future hypersurfaces $S$ play the same role in our
relativistic formalism as the final time coordinate, $t=T$, does in
the non-relativistic case.   

As in the non-relativistic case, we may assume that the relevant
fields are naturally divided into two (or more) classes.   
Our reason for employing this construction in the non-relativistic
case is the unphysical nature of non-relativistic propagators,
which imply that propagations from a single space point $x$ to 
any other point $y$ in any time $t$ are equally probable.  

Since relativistic propagators encode the causal structure 
of the underlying space-time, and tend rapidly to zero outside
the future light cone, it seems to us an open question -- which
depends on the details of the fields and their interactions and
the underlying assumptions about the initial state -- whether 
dividing the particles up into classes is necessarily required
in the case of relativistic field theory.   Our relativistic constructions
could be considered for a single class of particles, in which
case the stress-energy expectations at intermediate points would
be defined by post-selecting on the total final hypersurface
mass-energy density operator.  We intend to explore this 
possibility further in future.

We discuss here the case of two classes, $j=1$ and $2$, analogous to
the discussion given explicitly in the non-relativistic case above. 
Our definitions can easily be applied to the case of a single class.
Both the relativistic and non-relativistic 
definitions can also easily be extended to other post-selection rules, 
involving more than two classes; we discuss these possibilities 
in the next section.  

We use the following 
natural generalization of the final time measurements of mass density in our
non-relativistic models. 
For any given smooth hypersurface
$S$ in the future of the initial hypersurface $S_0$, 
we consider the effect of joint measurements of the class $1$
and $2$ local mass-energy
density operators $T^{(j)}_S (x) = T^{(j)}_{\mu \nu} (x) n_{\mu}
n_{\nu}$, 
carried out at 
each point $x \in S$, where $n_{\mu}$ is the forward-pointing timelike
unit $4$-vector orthogonal to the tangent plane of $S$ at $x$.  
We assume here the two mass-density operators commute.

This gives us a probability distribution on possible
mass-energy distributions $t^{(j)}_S (x)$ on $S$.  
Conditioned on any given outcome of the $t^{(j)}_S (x)$, we 
wish to calculate expectation 
values of the stress-energy tensors for the field
classes, $\langle T^{(\bar{j})}_{\mu \nu} (y) \rangle$,
at each point $y$ between $S_0$ and $S$.   

We again define these expectation values using the Aharonov-Bergmann-Lebowitz
formalism.   As in the non-relativistic case, we need to take
appropriate limits.  Again, because we have a post-selected final
outcome, the expectation value $\langle T^{(\bar{j})}_{\mu \nu} (y)
\rangle$ depends on the other commuting observables
that we consider as jointly measured.  
Because we no longer have an absolute time
coordinate, we need to define the relevant
measurement more carefully.

For any point $y$ in the future of $S_0$, define the {\it effective past
boundary} $\Lambda (y)$ of $y$ in our model to be 
$\Lambda_0 (y ) \cup S_0 (y ) $, where $\Lambda_0 (y)$ is the set
of points in the lightlike past of $y$ and the future of $S_0$,
and $S_0(y)$ is the set of points in $S_0$ not in the past light
cone of $y$.   Let $ \{ S_i (y ) \}$ be a sequence of smooth
spacelike hypersurfaces that include $y$ such that 
$$
\lim_{ i \rightarrow \infty} S_i (y ) = \Lambda (y) \, . 
$$
Consider a joint measurement of $T^{(\bar{j})}_{\mu \nu} (y)$ and 
of $T^{(\bar{j})}_{S_i (y)} (x)$ for all $x \in S_i (y)$ other than $ y$. 
Given the initial state $\ket{\psi_0}$ on $S_0$ and the
relevant post-conditioned final measurement
outcomes $t^{(j)}_S (x)$ on $S$, the ABL rule gives a value 
$$
\langle T^{(\bar{j})}_{\mu \nu} (y) \rangle_{S_i (y)}
$$
for the pre- and post-selected stress energy tensor expectation
value which, as our notation suggests, may in general depend on 
$S_i (y)$.   It is important to note 
that, as in the non-relativistic case, this 
expectation value depends on the full specification of the
measurement.  To apply the ABL rule, we need to include
all possible outcomes of all the measurements of $T^{(\bar{j})}_{S_i (y)} (x)$.  
We comment further on this later. 

Finally, we define 
$$
\langle T^{(\bar{j})}_{\mu \nu} (y) \rangle = 
\lim_{ i \rightarrow \infty } \langle T^{(\bar{j})}_{\mu \nu} (y) \rangle_{S_i
  (y)} \, , 
$$
assuming both that this limit exists and that it is independent of the 
chosen limit sequence $\{ S_i (y ) \}$. 

For a toy model in which all of physics takes place between $S_0$ and
$S$, the two functions
$t^{(j)}_S (x) $ on $S$ define the particular real
world that was randomly selected.  The pre- and post-selected
expectation values 
$\langle T^{(\bar{j})}_{\mu \nu} (y) \rangle$, for $y$ between $S_0$
and $S$
and $j=1$ and $2$, 
are the beables corresponding to the given real world, and define
physical reality between $S_0$ and $S$ in our model.  

We then consider the asymptotic limit in which $S$ tends to the
infinite future of $S_0$.   Suppose that $S_1$ is some fixed
hypersurface in the future of $S_0$. 
Let $C^{(k)}_{S_1}$ (for $k=1$ and $2$)
be any coarse-grained subsets of the sets of continuous
tensor functions $\{ t^{(k)}_{\mu \nu} (x) : x \in R^4 ,  S_0 < x < S_1 \}$,
where the notation $S_0 < x < S_1$ means that $x$ lies in the future
of some point in $S_0$ and the past of some point in $S_1$.  
Let
$$
\Prob_{S} ( C^{(1)}_{S_1} , C^{(2)}_{S_1} )
$$
be the probability that 
$\{ t^{(k)}_{\mu \nu} (x) : x \in R^4 ,  S_0 < x < S_1 \}$
belongs to $C^{(k)}_{S_1}$, for $k=1$ and $2$,
given our constructed probability density
function on the set of possible functions $T^{(j)}(x): S \rightarrow R $. 
Then, assuming that 
$$
\Prob_{\infty} (C^{(1)}_{S_1} , C^{(2)}_{S_1} ) = \lim_{S \rightarrow \infty} ( 
\Prob_{S} ( C^{(1)}_{S_1} , C^{(2)}_{S_1}  ) )
$$
exists, we define this to be the probability that reality between
$S_0$ and $S_1$ is described
by time-evolving mass distributions
belonging to $C^{(k)}_{S_1}$ for $k=1$ and $2$. 
This completes our proposed description of reality in this relativistic
model.  

Note that an interesting alternative model can be defined
by using the effective future boundary (defined analogously)
in place of the effective
past boundary. Again, this requires defining the expectation
value at $x$ via a limit using measurements on spacelike
hypersurfaces that tend to the effective future boundary of $x$. 
  
Another interesting possibility to explore, in discrete finite 
lattice models of space-time, would be to define the expectation value at
$x$ via a measurement on the first (or last) spacelike hypersurface
through $x$ from a foliation defined by stochastic
forward time evolution from $S_0$, as in \cite{dowker2004spontaneous}.

\section{Discussion} 

We have described a new way of defining a mathematically precise
description of physical reality that is not only consistent with
standard non-relativistic
quantum mechanics, but also involves only familiar quantities
that are simply defined within the theory, namely
expectation values of mass density. 
In contrast to Everettian ideas, 
this solution to the reality problem describes a randomly chosen single
real physical real world, selected from a well-defined probability
distribution in an entirely standard and 
unproblematic way. 
In contrast to de Broglie-Bohm theory, we believe our solution
will appear mathematically natural to anyone familiar 
with quantum theory.  
In contrast to dynamical collapse models, our solution requires
no change to quantum dynamics.  

We have also extended this to a Lorentz covariant solution of the
quantum reality problem for quantum field theory in Minkowski
space.   As in the non-relativistic case, this solution requires
assumptions about the asymptotic behaviour of solutions to quantum
dynamics given a realistic unitary evolution law and initial conditions.  

Our asymptotic assumptions can be tested directly in reasonably
complex models in the non-relativistic case, with the caveat
we noted above: to be reasonable tests, such models need to 
include assumptions that reflect the underlying field-theoretic
and cosmological intuitions.  Relativistic 
quantum field theory is not itself rigorously enough developed
to allow either our asymptotic assumptions, or the beable 
configurations they are intended to define, to be directly
calculated for complex systems.  Our solution to the reality
problem in this case thus involves formal definitions.
It could, however, still be tested in hybrid toy models in
which, for example, the asymptotic early and late time
states are taken to have fixed finite particle number.

Of course, {\it no} proposal
for solving the reality problem in relativistic quantum field
theory can be fully rigorously tested, 
given our presently limited understanding of the latter.
Ar present, the best one can hope for is
to show that there is a route to a solution
with no evident conceptual obstacles, and this we claim to have achieved.
Our proposal's ability to reproduce quasiclassical physics, and
the limiting behaviours it requires, can be tested in toy models. 

Relativistic quantum theory has been, purportedly,
one of our two fundamental theories of nature,
and may yet subsume the other -- general relativity -- in some
future quantum theory of gravity.  And yet, to date, it has been
completely unclear whether it admits {\it any} conceptually
clear description of physical reality, or allows a conceptually
clear derivation of classical dynamics or other higher-level
theories.   This has left serious questions over its status as a 
fundamental theory -- in Bell's words, it has seemed to ``carry the
seeds of its own destruction'' -- and led Bell and many others
to suspect that these
problems can only be solved by a deeper theory with different 
dynamics and experimental predictions.      
Replacing these fundamental conceptual problems by technical questions
about asymptotic behaviour -- in a theory that has in any case always
been understood to have deep unresolved technical questions -- seems
to us a considerable advance.    

We do not know for certain whether some appropriately
further extended version of our asymptotic assumptions
holds true in realistic cosmologies that include a theory of gravity,
or, a fortiori, whether it holds true in our universe. 
However, the essential idea that final states are asymptotically
well-defined superpositions of states of different mass density
configurations is at least consistent with some standard cosmological
pictures.  It is also consistent with the standard intuition that quantum field
theory should be understood as describing processes
from which asymptotically well-defined particle states emerge.   

Modulo these caveats, we believe our solution method is currently 
the most promising way of obtaining a physically sensible description
of a single quasiclassical world 
consistent with quantum theory and special relativity, and 
plausibly consistent with gravity and cosmology.   

The method could, of course, be applied to other physical quantities,
and so our solution is not unique.  
For example, probability or charge densities
could be used instead of mass densities in the non-relativistic case.
In the relativistic case, the final measurements could be of $J_{\mu} n_{\mu}$,
and the pre- and post-selected 
expectation values of the electromagnetic $4$-current $J_{\mu}$
could be used instead of that of the stress-energy tensor to define
the real beables.   

Other possibilities could also be considered.  
Nonetheless, relatively few options seem particularly natural, and
among these, mass density (in the non-relativistic case) 
and the stress-energy tensor (in the relativistic case) 
seem to us the most natural.   
A strong additional motivation for focussing on these options is 
that they suggest new ideas to explore in unifying quantum theory
with gravity: we discuss these further below. 

There are also various ways in which the particles (in the
non-relativistic case) or fields (in the relativistic case)
could naturally be divided into two or more classes, and 
various post-selection rules that could be considered.
For example, one might take the classes to be 
bosons and fermions, or massive and massless particles
(or, more speculatively, matter and gravitational fields, or ordinary
and dark matter, in the appropriate contexts). 
Which choice(s) of classes are most natural depends on the 
Hamiltonian and the asymptotic form of the final state
(and thus also on the initial state). 
Any given physical theory in which these are specified
should allow relatively few options that seem particularly natural.

In the variants of our model in which several classes are 
considered, one natural rule for 
defining the class $j$ mass-densities is to post-select
on the final outcomes for all the other classes; again,
other rules could be considered.
Once again, though, relatively few options for choosing 
classes, or post-selection rules, are likely to 
seem particularly natural, given a specific theory.

One other variant of our model that is worth noting is that
in which there are two classes of particles or fields, but
only the mass density (respectively mass-energy density) for
one of them defines beables.  While this seems less natural
if the ultimate aim is to couple the mass(-energy) beables 
directly to a quasiclassical gravitational field, it 
seems adequate as a solution to the quantum
reality problem {\it per se}: a quasiclassical picture of reality
can seemingly be described padequately in terms of the mass-energy densities
of fermions, or massive particles, for example.   
 
\subsection{Implications for earlier approaches to the quantum reality
  problem}

Although admittedly incomplete, this work 
raises, in our view, new questions about previous approaches to the reality
problem.   For example: 

Why resort to de Broglie-Bohm theory,
with its inelegant combination of particle-like trajectories 
guided by an evolving quantum wave function, if a solution to
the non-relativistic quantum reality problem exists that uses 
only simple quantities that arise naturally in quantum theory?   
The case against de Broglie-Bohm theory seems all the
stronger when we consider the relativistic reality problem,
and the fundamental conceptual problems that arise when one tries 
to define any fully Lorentz covariant version of de Broglie-Bohm
field theory. 

Similarly, why resort to many-worlds ideas, if there is a simple one-world 
solution to the reality problem?   Why try to deal with the problem
of the appearance of quasiclassicality in many-worlds quantum theory,
and the necessary imprecision in defining the branching worlds, when
we can give a simple picture with a single, precisely defined 
quasiclassical world?   And why struggle with what seems to many 
(e.g. \cite{kentoneworld,albert,price}) 
the hopeless task of trying to make sense of probability in a 
deterministic many-worlds theory, if a straightforwardly
probabilistic one world description is available?    

Moreover, if there is a reasonably natural way of solving the reality problem
within standard quantum mechanics, do we need to consider collapse
models, with their ad hoc assumptions and extra parameters?  The
question seems even more apt given that this solution
also extends naturally to relativistic quantum theory and -- while 
admittedly not rigorously defined in this context -- still appears
to pose fewer
technical or conceptual problems than attempts at relativistic
generalizations of collapse models.  

Bell said \cite{bell2001there} of
Ghirardi-Rimini-Weber's original discrete dynamical 
collapse model \cite{ghirardi1986unified}: ``I
am particularly struck by the fact that the model is as Lorentz
invariant as it could be in the nonrelativistic version. It takes away
the ground of my fear that any exact formulation of quantum mechanics
must conflict with fundamental Lorentz invariance.''
The ideas for a 
solution to the reality problem outlined in this paper take away
the ground of my own prior hunch that any exact Lorentz invariant
formulation of quantum theory must {\it necessarily}
alter the dynamical equations
(as the GRW theory and other dynamical collapse models do).  
Given the extraordinary beauty of both special relativity and quantum
theory this prompts the question: if we 
can solve the reality problem and retain both theories intact, 
(why) would we want to consider alternatives that break one or the other?    

\subsection{Generalizations of quantum theory}

\subsubsection{1. Generalizations using beable guided quantum theory}

This last question has some real force -- in particular, if the
ideas outlined here work, then 
the case for dynamical collapse models does seem weakened.
However, it is by no means purely rhetorical.  
There {\it are} still good reasons for continuing to explore generalizations
of quantum theory, and indeed the solutions to the quantum reality 
problem described above also suggest intriguing
new directions for such exploration.  

One entirely uncontroversial motivation is that, however beautiful quantum
theory appears, and even if the reality problem and all other 
conceptual and technical issues can be resolved within standard
quantum theory, it still may turn out not to be the final theory
of nature.  We want to test our best current theories as strongly
as we can.  To do this, we need alternatives against which to 
test it -- and ideally, we would like parametrised classes of
alternatives to quantify the extent to which it has been tested. 
Such alternatives need not be as beautiful or compelling as
our best theory -- indeed, almost by definition, they will not
be.  They can still serve a valuable role as foils, or, to be
open-minded about it, as ways of pointing out domains in which
our best theory might in fact break down -- even if not necessarily
quite for the reasons those alternative theories suggest. 

There is, though, also a strong case for taking generalizations of quantum 
theory seriously on their own terms \cite{kent2013beable}. 
Any solution to the
quantum reality problem defines a probability distribution
on configurations of beables.  It is perfectly logically
consistent for this distribution to be defined only by the 
initial conditions and quantum dynamics, as our solutions
are.  But there is, arguably, something oddly epiphenomenal
about the status of the beables in such a theory.  On the one hand,
they are the building blocks of physical reality.  On the
other hand, they seem to play a mathematically secondary
role to that of the evolving quantum state.  It determines
their probability distribution, while they have no effect on it.  

Of course, it could be 
that nature is described this way.  It is hard to know just
how much weight to put on the intuition that physically crucial quantities
in a fundamental theory should play a more central role
in the mathematics \footnote{It would be interesting to see
a Bayesian analysis of this, based on how well scientists' intuitions 
about fundamental questions have fared over the centuries.
The fact that we have made a lot of progress, some of it guided
by intuition as well as experiment, suggests our intuitions are
not worthless.   On the other hand, the library shelves littered with confident 
but misguided pronouncements about the way nature must be, 
and the many surprises in physics
that no one would or could have predicted, suggest that we 
may tend to overrate intuition.  My hunch is that such an analysis
would show, inter alia, that the loudest pronouncements do not tend to
be the most reliable.}.   Still, the intuition is there.
It also motivates a class of generalizations of quantum 
theory, which moreover suggest a new way of thinking about
quantum theory and gravity.  So, whether or not the
underlying intuition is fundamentally right,
it suggests potentially valuable new directions for theoretical
physics.   

Recall that, given initial conditions
and dynamics, our solution defines a probability distribution
on configurations of beables -- in the form of space and time-dependent 
mass density or stress-energy tensor expectation values --
that define reality.   Any such construction can be 
generalized by taking the probability distribution to 
depend not only on the initial conditions and dynamics
but also on the beable configuration itself \cite{kent2013beable}. 
To give just one example among countless possibilities, the probability
of configurations could be enhanced or suppressed
depending on some global measure of its uniformity
over time.  Different generalizations can be 
obtained from versions of our solution involving charge density or
other quantities.    

The choice of stress-energy tensor expectation values as
beables for relativistic quantum field theory
is particularly suggestive if we allow the probability
distribution for configurations of these beables to depend 
on laws not defined only by the evolving quantum state.  
This suggests the thought that it might be possible to
unify gravity and quantum theory via 
probabilistic quasiclassical laws that couple the background geometry
directly to the quasiclassical beables defining a matter 
distribution, without necessarily requiring
any quantised gravitational field.

The idea here is not to restrict to defining versions of semi-classical
gravity by equations of the form
$$
G_{\mu \nu} + g_{\mu \nu} \Lambda = 8 \pi \langle T_{\mu \nu} \rangle
\, ,
$$
where the expectation value on the right hand side is defined
by one of the recipes given above (summing over the expectation
values for the classes if there are two or more classes). 
Such equations need not generally be everywhere consistent. 
Instead, the goal is to extend the probability distributions defined
above on beable configurations, represented by $\langle T_{\mu \nu}
\rangle$,
to joint probability distributions on Riemannian manifolds and 
tensor fields defined on such manifolds, with the property that
the quasiclassical Einstein equations emerge as approximately
valid in appropriate domains.   
We intend to explore this further in future work.  

\subsection{Relation to previous work}

The solution to the reality problem outlined here uses 
the strategy of inferring finite time beables from asymptotic
behaviour, and many of the other ideas,
set out in Ref. \cite{akrealworld}, but is
simpler than the proposals made in that paper. 
While the simplicity of the solution given here is particularly
appealing, and its suggestion of a relationship to gravity
particularly intriguing,
those earlier proposals still
remain potentially interesting alternatives, in our
view.  Both the solution proposed here and those in Ref. \cite{akrealworld}
have features in common with
other earlier ideas in the literature.   

The fundamental significance of the quantum reality problem
and the possibility of finding a mathematical solution 
was perhaps first realised by de Broglie and Bohm \cite{debroglie,bohm}.
The concept of beable is due to 
Bell \cite{bell1976theory,bell1987beables}, who
also illustrated the variety of possible types of beable solution
to the reality problem and focussed attention on
the Lorentzian quantum reality problem \cite{bell2001there}.   

Aharonov and collaborators \cite{aharonov1964time,
aharonov2008quantum,
vaidman2012probability} have 
long stressed the value of considering both initial and final 
states in order to illuminate the properties
of and interpret quantum theory from various perspectives.
Related ideas were previously 
considered by Watanabe \cite{RevModPhys.27.179} 
Suggestions for interpretations of 
quantum theory in terms of 
initial and final states have been made by 
Davidon \cite{davidon} and Aharonov and Gruss \cite{aharonov2005two}. 
From the perspective adopted here, one major limitation of these 
latter ideas is their reliance on intuitive definitions of measurement
and classicality, which are unsatisfactorily imprecise in any setting
and especially problematic in the context of cosmology; see 
\cite{bell2004speakable,akrealworld} for further discussion.    
Another fundamental problem, from the perspective of those looking for a new
one-world solution to the reality problem, 
is a super-Everettian ontology that includes the 
evolving forward wave function.  

The idea of defining cosmological models with independent
initial and final boundary conditions, using a decoherent
histories version of the ABL rule, was discussed by
Gell-Mann and Hartle \cite{gell1990quantum}.   
The possibility of defining cosmological models and other
generalizations of the quantum theory of closed systems 
by going beyond boundary conditions, and 
considering a sequence of constraints on the system's evolution,
was proposed in Ref. \cite{kent1998beyond} and developed and
discussed further in Ref. \cite{kent2013beable}.   
The potential uses of environmental records in making 
sense of quantum theory and quantum cosmology have been
stressed by Zurek and 
collaborators \cite{zurek1993preferred,blume2006quantum,zurek2009quantum} 
and by Gell-Mann and Hartle \cite{gell1990quantum}, among others.  

While all of these contributions have
been influential and relevant to our discussion, 
none of the above authors have
proposed a mathematically precise solution to the Lorentzian quantum
reality problem in the sense defined by Bell and considered here.  

Mass density beable ontologies were first proposed 
for non-relativistic collapse models by Pearle and Squires
(\cite{PhysRevLett.73.1}; see also \cite{ghirardi1995describing}).  
An extension of these ontologies to relativistic collapse models using 
constructions previously defined   
in \cite{kent2005nonlinearity,akcausalqt}
was proposed in \cite{bedingham2011relativistic}.  
These proposals apply to generalizations of quantum theory
rather than to quantum theory itself.  
We presently see our relativistic solution as more natural and 
see the path to rigorizing it as having  
fewer (although still considerable) technical obstacles.

We also see our proposed solutions as calling into question
part of the original motivation for dynamical collapse models. 
It should be noted, though, that 
one possible motivation for dynamical collapse models is the
desire for a theory that effectively {\it ensures} that 
macroscopic superpositions essentially never take place,
even in hypothetical future experiments in which technology
allows us either to isolate macroscopic systems for long
times, or to control their environments, in such a way
that quantum theory would predict a genuine macroscopic
superposition and an ensuing quantum interference pattern.
Our solutions suggest an ontology in which all significant
components of the macroscopic superposition have corresponding
beable trails in such experiments.   This does not seem evidently
problematic: there is no logical inconsistency in such a 
description, nor any contradiction with experiment or observation
to date.  Still, those who prefer the hunch that nature abhors
a macroscopic superposition will prefer collapse models or others
with this feature. 

Of course, all these various questions certainly deserve 
further analysis.  We also
wish to stress that, whether or not they ultimately prove
relevant to nature, dynamical collapse models remain in our view a landmark
intellectual achievement in the development of work on the 
quantum reality problem, and that there remains a 
scientific case for exploring 
them simply because they are testable generalizations of quantum
theory.

\section{Acknowledgements}
This work was partially supported by a Leverhulme Research Fellowship,
a grant from the John Templeton Foundation, and by Perimeter Institute
for Theoretical Physics. Research at Perimeter Institute is supported
by the Government of Canada through Industry Canada and by the
Province of Ontario through the Ministry of Research and Innovation.
I am very grateful to Philip Pearle for pointing out a problem
in an earlier version, and to Fay Dowker, Philip Pearle and John
Taylor for other very helpful comments.

\section {Appendix 1: 
Beable models based on other definitions}

In this appendix we consider alternative strategies for defining
beables, applicable to both non-relativistic and relativistic models.
We make the same assumptions as previously about
the initial state and asymptotically defined final conditions.

\subsection{Expectation values defined by measurements at 
a single point }

In our non-relativistic model, we defined 
the beable at the point $(x,t)$ 
in terms of an expectation value 
$ \langle \rho (x,t) \rangle $.
This expectation value was defined via the ABL rule 
for a joint measurement of mass density 
at all points $y$ on the surface of constant time $t$. 
We could, instead, have defined an expectation value,
which we denote $ \langle \rho(x,t) \rangle_{\rm x}$,
by applying the ABL rule for a single measurement of 
the operator $ \rho_(x,t) $.  

As Aharonov and Vaidman's box 
``paradoxes'' \cite{aharonov1991complete} illustrate, there 
are final states for which this would give a different
beable distribution, with somewhat counterintuitive
properties.    The beables for a single particle in
a three box example would suggest that its entire 
mass was simultaneously in two distinct regions (with
some further nonzero mass density expectation in a third).
Mass would thus not generally be conserved at the beable
level.  

This is aesthetically worrying, and even more troubling
if one hopes that the mass density beables play a significant
role in combining quantum theory and gravity (a possibility
we consider further below).   
Nonetheless, we should note that such a description is 
not logically inconsistent, and 
it is not immediately evident to us that it is incapable of 
reproducing quasiclassical physics.  In principle, a beable version
of quantum theory might give a counter-intuitive picture
of reality in microscopic experiments (or indeed in 
macroscopic experiments that are to date unperformed)
and still allow the derivation of the correct
higher-level quasiclassical laws in the right regime. 

Similar comments apply to the relativistic models. 
We could define a stress-energy  tensor beable
$\langle T_{\mu \nu} (y) \rangle_y$ as the ABL expectation
value for a measurement of $T_{\mu \nu} (y)$ alone.  
This has what might be seen as the advantage of dispensing
with a definition based on a limit of spacelike hypersurfaces that
approximate and tend to the past light cone.   
It has the same counter-intuitive features as its non-relativistic
counterpart in its description of three-box experiments, however. 

We should also note that, while the beable models we defined earlier
give more intuitively sensible descriptions of three-box experiments,
we would not expect them to agree with all prior intuitions in their
descriptions of every closed quantum system.  
For example, the stress-energy tensor beable $\langle T_{\mu \nu} (y)
\rangle$ need not generally satisfy conservation laws everywhere, 
and cannot be directly used as a source in Einstein's equations 
for a semiclassical treatment of gravity.   
(Indeed, no quasiclassical derivation of Einstein's equations
involving macroscopic quantum experiments can be valid everywhere
 \cite{kent2013might}.) 

In summary, models in which the beables are defined
by single point expectation values have decidedly odd features.
While these make such models seem presently less attractive, we 
do not feel they presently exclude them.  
More analysis of the relationship between the beable 
distributions in these models and quasiclassical
variables is needed.    

\subsection{Comments on weak operator values and beables}  

Another candidate beable in our relativistic models could be the 
modulus of a generalised form of the so-called
weak value \cite{aharonov1991complete,aharonov2005two} 
of the stress-energy tensor
$$
 {\cal A}_{\mu \nu} (x)_w  =  \left| \left( \frac{ 
\Tr( T_{\mu \nu} (x) { P_S }{T_{\mu
      \nu}(x)}{P_0})}{  \Tr ( P_S P_0 ) }  \right) \right|^{1/2}  \, .
$$ 
Here $P_S$ is the projection onto the space of states whose 
mass-energy distributions $t_s (x)$ agree with that randomly
selected on the final hypersurface $S$, $P_0 = \ket{ \psi_0 } \bra{
  \psi_ 0 }$ is the projection onto the initial state on $S_0$.
These are unitarily evolved backwards
and forwards, respectively, to any chosen spacelike hypersurface $S'$ 
through $x$, and the inner products are calculated on $S'$: we suppress
these details in our notation. 

Similar comments apply to this and other quantities derived
from weak expectation values or decoherence functions.   
Such quantities behave similarly to the single point expectation
values in three-box experiments, so that a single particle
would appear in the beable description to be located in more
than one box, and have other peculiar properties (see e.g. 
\cite{svensson2013quantum}).   
This does not logically exclude them as candidate beables capable
of describing quasiclassical physics, but motivates further careful
analysis.   

\section{Appendix 2: Further generalizations of quantum theory}

\subsection{Generalizations by taking finite limit parameters}

Consider again our non-relativistic model, in which a mass
distribution is obtained at final time $T$, and then used to
define mass density distributions at times $0 < t < T$. 
These expressions are calculated by using 
projectors $P^{\Delta}_{\rho_f}$ onto 
the set of states with final mass distribution in a neighbourhood
$\Delta$ of $\rho_f (x, T)$ and taking the limit as the size of 
$\Delta$ tends to zero. 
Finally, we take the limit $T \rightarrow \infty$.  
These limits are 
intended to reproduce a quasiclassical reality consistent
with standard quantum theory in realistic models.   

To produce generalisations of quantum theory, we can take
$T$, and if we wish also $\Delta$, to be fixed finite parameters.      
Intuitively, if $T$ is large compared to the duration of a 
quantum experiment, this gives predictions almost indistinguishable
from those of standard quantum theory for that experiment.  

Of course, taking the finite $T$ version of the model literally
suggests that reality ceases after time $T$ has elapsed -- even
though (on a standard reading) 
the quantum dynamics may continue to be eventful long afterward. 
Our recommended
attitude to this is not to take the model literally on this point.
A commonly held view of 
dynamical collapse models is that 
although the mathematical details of their collapse mechanisms
are ad hoc, and it is hard to believe that either they or 
the associated ontologies are fundamentally correct, the models 
are nonetheless interesting generalizations of
quantum theory.   They make an intellectually significant point --
altering quantum dynamics somewhat radically alters the
ontology and gives new solutions
to the quantum reality problem -- 
and also point to interesting experimental tests.
A model does not need to be completely right in order to point
in a direction that is theoretically or experimentally fruitful to explore. 
Similarly, finite $T$ and $\Delta$ versions of our models show that
altering quantum theory gives a well-defined
realist ontology, without any assumption about the asymptotic
dynamics, and in a way that affects the predictions for quasiclassical
dynamics and experiment so subtly as to be essentially 
undetectable.\footnote{ 
It also gives a different way of parametrizing how well quantum
theory is tested in any given experiment. Note though that no foreseeable 
experiment is going to test the parameter range in which $T$ is
large in human lifetimes, and it is hard to take the models seriously
if $T$ is short compared to a human lifetime.} 

Another possible approach to finite $T$ models would be to construct versions
in which ``final measurements'' are made repeatedly on
timescales of order $T$, and the chosen reality depends on a 
sequence of final measurement outcomes in such a way that it evolves
smoothly.   At first sight, such models look mathematically rather ad
hoc, since one can imagine many recipes of this type, none of which
seems particularly natural.  They also look likely to have 
physically peculiar consequences, in which reality is something like a real
superposition (with time-evolving weights) of a sequence of
independently randomly chosen realities defined by measurements at
times $\approx T , \approx 2T , \ldots $.   Perhaps, though, there 
is scope to construct more natural models by variations on these
ideas: we leave this for future exploration.   

\subsection{Comments on relativistic generalizations} 

One might also investigate ``finite proper time''
generalizations of our relativistic models.
Given initial data on a spacelike hypersurface $S$, and a time
parameter $\tau$, we can define a finite version of the models
given any Lorentz covariant rule that produces a final
hypersurface $S'$ that depends (stochastically or deterministically)
on $S$ and $\tau$, with the property that $S'$ is always in the 
future of $S$ and that the maximum proper time between points on
$S$ and $S'$ is a function of $\tau$.   
Such rules can be naturally defined in 
finite lattice models \cite{dowker2004spontaneous}.
It would be interesting to explore continuous versions in Minkowski space, 
or indeed analogous rules in quantum gravity using natural
definitions of cosmological time \cite{york1972role}.   

Again, a literal reading of such models
suggests that reality exists only between $S$ and $S'$;
again, our preferred attitude would be not to take any such
model so literally.
Another feature is that the
choice of hypersurfaces $S, S'$ suggests some sort of preferred
coordinate choice, even if the rules relating $S'$ to $S$ are 
Lorentz covariant.  This may not necessarily be problematic -- after all,
standard general relativistic cosmological models
can also include both preferred proper time coordinates and associated
spacelike hypersurfaces -- but requires careful discussion.
Again, we leave these preliminary ideas for future exploration.

\section*{References}

\bibliographystyle{plain}
\bibliography{reality}{}

\end{document}